\documentclass[aps,prb,twocolumn,groupedaddress,showpacs]{revtex4}

\usepackage{graphicx}
\begin{document}
\renewcommand{\textfraction}{0.15}
\renewcommand{\topfraction}{0.85}

\title{Effect of well-width on the electro-optical properties of a quantum well}
\author{Arjun Joshua}
\email{arjun@physics.iisc.ernet.in}
\author{V. Venkataraman}
\affiliation{Department of Physics, Indian Institute of Science\\
Bangalore, India 560012}

\date{\today}

\begin{abstract}
We record photoreflectance from Ge/GeSi modulation doped quantum
wells possessing $10^4$ V/cm perpendicular electric fields.
Qualitatively very different spectra are obtained from samples of
well-width 100 \AA\ and 250 \AA. Comparing the wavefunctions
calculated from an $8 \times 8$ \textbf{k.p} theory, we find that
while they remain confined in the narrower 100 \AA\ QW, the
electric field causes them to tunnel into the forbidden gap in the
250 \AA\ well. This implies that the samples should show a
transition from the quantum confined Franz-Keldysh effect to the
bulk-like Franz-Keldysh effect. Close to the band-edge where Franz-Keldysh effects are important, simulated photoreflectance spectra reproduce the essential features of the experiment, without any adjustable parameters.
\end{abstract}

\pacs{78.67.De, 78.40.Fy, 78.20.Jq}
\maketitle

\section{Introduction \label{intro}}

An electric field has the ability to drastically alter the optical absorption in a quantum well (QW). The field can be in the plane of the well or perpendicular to it's layers. Parallel fields cause changes in QW absorption that resemble bulk; the broadening of excitonic features and the onset of FK oscillations has been observed in InGaAs/GaAs by photocurrent.~\cite{schmeller} We shall instead consider the perpendicular field case, where the field lies normal to the plane of the QW. For narrow wells where the well width $L_z$ is less than the 3D exciton Bohr radius, excitonic effects dominate electro-absorption. This is the quantum confined Stark effect which has been studied by theory and experiment in materials where excitons are important, e.g., GaAs~\cite{millerc} and ZnSe.~\cite{merbach} However when $L_z$ is larger than the exciton size, but not large enough for quantum effects to be ignored, Coulombic effects can be neglected and QW absorption in an electric field is conventionally called the quantum confined Franz-Keldysh effect (QCFKE).~\cite{miller, millerd} This regime applies to room temperature electro-absorption in Ge or Si QWs, since excitonic effects in these materials are weak. As $L_z$ is increased further, quantum size effects become less important. Also the continuum of wavefunctions begins to tunnel into the band-gap, giving rise to the Franz-Keldysh effect (FKE).~\cite{tharma, aspnes1} This article is concerned with the $L_z$-dependent trend occuring in the optical properties of QWs from the QCFKE to the bulk-like FKE.

Experimental techniques that have been used to study the higher energy transitions in a QW are photocurrent and photoreflectance spectroscopy.  External contacts or modulation  doping are used to create an electric field in the sample. There have been several experiments where the increase of forbidden transitions with a perpendicular electric field has been observed, in QWs of  fixed $L_z$.~\cite{wang, collins, yamanaka} For perpendicular fields, Miller \textit{et al.}~\cite{miller} analytically treated the $L_z$-dependent transition from the QCFKE to the FKE, assuming infinite barriers and in a two-band model. They showed that the QCFKE in a GaAs QW approached bulk electroabsorption when the critical thickness was only about 300 \AA. Under identical assumptions but applying a scaled Fourier Transform to the absorption, Chen \textit{et al.}~\cite{chen} have recently suggested that the critical thickness in GaAs is in-fact 600 \AA\  instead of 300 \AA. A QCFKE to FKE transition has been observed by Keil \textit{et al.}~\cite{dohler} who have performed photoreflectance in GaAs/AlGaAs \textit{n-i-p-i} superlattices. The built-in field was changed by thermally generating carriers keeping $L_z$ constant at 300 \AA. No experiments have so far been reported in the literature where the trend from the QCFKE to the FKE has been studied by keeping the field constant and increasing the well width $L_z$, although this is the intuitive way to reach bulk behavior. It also permits easy comparison with the earlier theoretical results where QW electroabsorption was calculated as a function of well width.~\cite{miller,sanders} We present photoreflectance experiments on 100 \AA\  and 250 \AA\  modulation doped strained Ge single QWs, having $\approx10^4$ V/cm built-in electric field. In these samples excitons can be neglected at room temperature. Simulation of the observed spectra based on a 8-band  \textbf{k.p} calculation with no fitting parameters, demonstrates a trend from the QCFKE to the FKE as $L_z$ is increased.

Photoreflectance (PR) has proved useful in earlier reports of the characterization of undoped QWs in the Ge/GeSi system.~\cite{pearsall,rodrigues, yaguchi} Pearsall \textit{et al.}~\cite{pearsall} have measured strain and confinement shifts of the transition energies in GeSi multiple QWs. Using PR, average valence band-offsets in Ge/GeSi superlattices have been determined by Rodrigues \textit{et al.}~\cite{rodrigues} Yaguchi \textit{et al.}~\cite{yaguchi} have studied offsets in the conduction and valence band by characterizing Ge/GeSi strained barrier single QWs.

\section{Experimental Details \label{exp}}

\subsection{Photoreflectance}
The set-up is of the conventional kind. Radiation from a 100 W quartz tungsten halogen lamp was filtered through a 1/4 m Jobin-Yvon monochromator to serve as a probe ($\approx 10 \mu$W). It was then focused onto the sample by a lens at an incident angle of $45^{\circ}$. After reflecting off the sample, the probe was imaged onto a detector by another lens. The detector used was a Si photodiode for probe energies greater than 1.1 eV, and an InGaAs photodiode for lower energies.  A focusable 2.4 mW, 670 nm laser diode was mechanically chopped at $\approx$ 187 Hz and was normally incident on the sample so as to overlap with the probe spot. The spot size was 1$\times$1mm$^2$. Appropriate order-sorting filters and blocking filters prevented unwanted light from reaching the detector. The laser light was absorbed in the sample close to the surface, giving rise to free electrons and holes in this region. They served to modulate the electric field in the sample. The consequent modulation in probe reflectance was picked up by the detector and was feeded to a SRS830 lock-in amplifier. The dc and ac components (at the chopping frequency) of the amplifier signal give $R$ and $\Delta R$ respectively. They were acquired by a computer and digitally divided to obtain the photoreflectance signal $\Delta R/R$. All experiments were performed at room temperature, with a spectral resolution of $\approx$ 10 meV. The experimental range of our set-up was 0.7 eV to 1.6 eV.

\subsection{Sample}
\begin{figure}[bp]
\includegraphics[height=7cm, keepaspectratio, angle=-90]{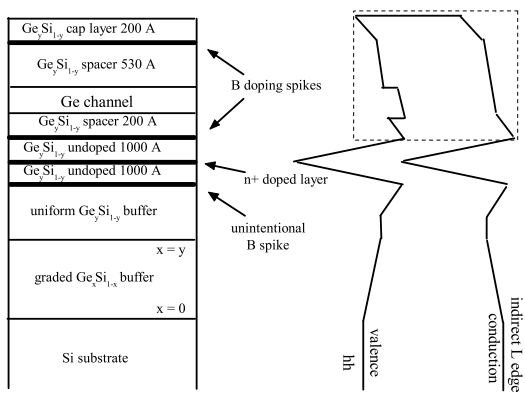}
\caption{\label{fig:strucband}Generic sample structure along with schematic band-diagram. Bold lines in the structure, on the left, represent doped layers. The corresponding band diagram shows the highest valence band (hh) and the lowest (indirect) conduction band .}
\end{figure}

The sample structure and corresponding schematic band-diagram for the valence and lowest conduction band which is indirect (L-point) are shown in Fig.~\ref{fig:strucband}. The samples were grown by molecular-beam epitaxy on Si (100) substrates.~\cite{xie1,xie2} A 10 $\mu$m thick compositionally  graded Ge$_{x}$Si$_{1-x}$ buffer is grown first, with $x$ varying from 0 to $y$ ($y$ = 0.7 or 0.8). In order to improve the yield, during buffer growth the samples were taken out and cut into quarters leading to some unintentional boron doping. An $n^+$ layer (shown as a large spike in the band-diagram) was introduced during further growth to isolate the QW from the substrate and to provide a reference point for the energy bands. After subsequent boron delta-doping, an unstrained 200 \AA\ Ge$_{y}$Si$_{1-y}$ spacer layer is  deposited. A Ge layer is grown next, compressively strained to the lattice constant of the underlying Ge$_{y}$Si$_{1-y}$ alloy. A second Ge$_{y}$Si$_{1-y}$ spacer of thickness 530 \AA\ separates the Ge layer from yet another boron delta-doped region. The delta-doping in both the spacer layers is $\approx 10^{12}$ cm$^{-2}$. The structure is finally capped by 200 \AA\ of Ge$_{y}$Si$_{1-y}$. Two different samples were studied by PR. The 100 \AA\ Ge layer sample (\#411) had Ge$_{0.7}$Si$_{0.3}$ alloy layers on a $n^+$-Si substrate. The other sample (\#348) had a 250 \AA\ Ge layer with Ge$_{0.8}$Si$_{0.2}$ layers on a p-Si substrate.

\begin{figure}[tp]
\includegraphics[height=5.5cm, angle=0]{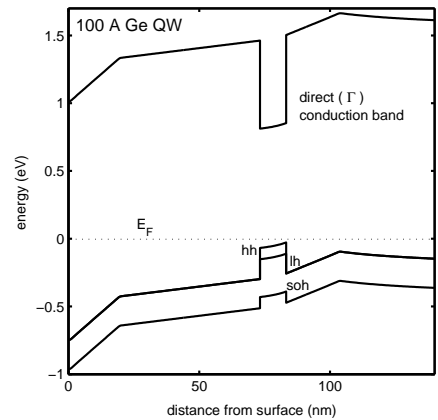}
\caption{\label{fig:calcband411}100 \AA\ Ge/Ge$_{0.7}$Si$_{0.3}$ QW. The calculated bands are shown for the boxed region of Fig.~\ref{fig:strucband}, with the lowest direct replacing the indirect conduction edge. The valence band degeneracy is shown resolved by strain, into the heavy and light hole bands. The split-off hole band is also shown. The electric field in the well (3.9$\times$10$^4$ V/cm) arises from the surface depletion layer and the boron $\delta$-doping. The laser is off. The dotted line refers to the equilibrium Fermi energy $E_F$.}
\end{figure}

\begin{figure}[tp]
\includegraphics[height=5.5cm, angle=0]{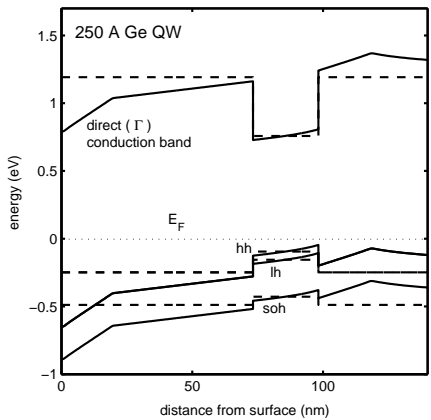}
\caption{\label{fig:calcband348}250 \AA\ Ge/Ge$_{0.8}$Si$_{0.2}$ QW. The field inside the well is 3$\times$10$^4$ V/cm. Same comments apply as for Fig.~\ref{fig:calcband411}. Bold lines indicate the laser is off and dashed lines indicate the laser is on, neutralizing the field.}
\end{figure}

In the lowest conduction band there is negligible confinement (Fig.~\ref{fig:strucband}). The QW therefore contains predominantly holes injected by the boron delta-doping. The measured carrier concentration inside the 100 \AA\ Ge QW~\cite{madhavijap} is 7$\times$10$^{11}$ cm$^{-2}$, while that for the 250 \AA\ Ge QW~\cite{madhaviprb} is 8.6$\times$10$^{11}$ cm$^{-2}$. At room temperature, this corresponds to a Fermi energy which lies just outside the well (dotted lines in Figs.~\ref{fig:calcband411} and \ref{fig:calcband348}), and thus band-filling effects are not expected to be significant.

Though Ge and Si are indirect gap materials, it is their direct edges that contribute to the photoreflectance signal. The direct gap of Ge (0.798 eV) at the $\Gamma$ point, is less than that of Si (4.00 eV) (see Table.~\ref{table1}). After band alignment, Ge serves as a QW in both the direct conduction and in the valence band. The barrier is formed by the Ge$_{y}$Si$_{1-y}$ layers. Figs.~\ref{fig:calcband411} and~\ref{fig:calcband348} shows the calculated band-diagram of the region comprising the surface depletion layer, the alloy barriers, the B-delta doping and the Ge QW. This active region is about 120 nm thick. We expect the $n^+$-doping along with the unintentional B-doping to have a negligible combined effect on the quantized energies in the Ge layer and we have not included them in the calculation. The $\approx10^4$ V/cm calculated field in the QW is thus attributed to the boron delta doping and the depletion-region at the surface. When the laser is shone on the sample, it generates holes and electrons by band-to-band absorption with a sheet density of $\approx 10^{12}$ cm$^{-2}$. This is comparable to the carrier density created by modulation doping. Thus it seems reasonable to assume that the laser excitation neutralizes band-bending leading to flat bands, as indicated schematically by the dashed lines in Fig.~\ref{fig:calcband348}.

\section{Experimental results}

\begin{figure}
\includegraphics[width=8cm, angle=0]{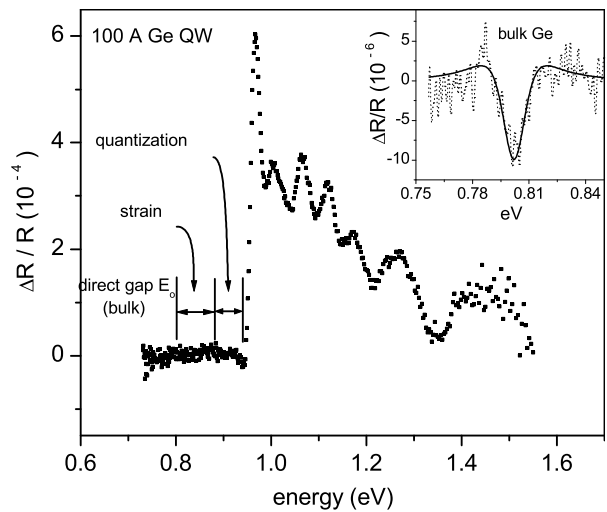}
\caption{\label{fig:411ge_inset}PR from 100 \AA\ QW. The horizontal arrows indicate the contribution of strain and quantization, from a finite square well model, in shifting the band-edge from the bulk value. The inset shows PR from bulk Ge.}
\end{figure}

As an experimental check on our setup, PR was recorded from bulk Ge and fitted to obtain the direct edge. The value obtained was compared with literature. The inset in Fig.~\ref{fig:411ge_inset} shows the signal from a 1 $\mu$m thick Ge film on Si. The film was undoped and strain relaxed. For low electric fields $F$, PR from a 3D critical point is given by Aspnes' line-shape function~\cite{aspnes}
\begin{equation}
 \frac{\Delta R}{R}=Re\left[\frac{Ce^{i\theta}}{(E-E_o + i\Gamma)^{2.5}}\right], \label{prfit}
\end{equation}
where $C$ and $\theta$ are amplitude and phase factors, varying slowly with energy $E$. $E_o$ and $\Gamma$ are the critical energy and broadening respectively. The solid line in the inset depicts the fit of this line-shape to the data. Fitting gave $E_o=0.802$ eV, close to the accepted value of 0.798 eV (see Table~\ref{table1}) for the direct edge of bulk Ge. The corresponding value of $\Gamma$ is 14 meV, larger than the nominal resolution of 10 meV, since the monochromator slits were widened to record the weak signal.

\begin{figure}[htp]
\includegraphics[width=8cm, angle=0]{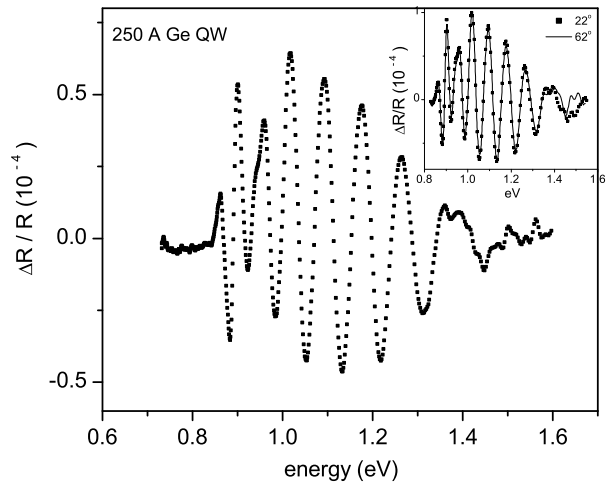}
\caption{\label{fig:348ang_inset}PR from 250 \AA\ QW at a default $45^{\circ}$ incident angle. The inset shows PR from the same sample at $22^{\circ}$ and $62^{\circ}$.}
\end{figure}

PR spectra from the 100 \AA\ and 250 \AA\ QWs are shown in Fig.~\ref{fig:411ge_inset} and Fig.~\ref{fig:348ang_inset}. The 100 \AA\ Ge/Ge$_{0.7}$Si$_{0.3}$ spectrum shows it's first peak at 0.95 eV. The effect of strain and quantization in causing a blue-shift compared to bulk Ge is indicated by arrows in Fig.~\ref{fig:411ge_inset}. In-plane biaxial compressive strain in the Ge layer causes tensile strain in the transverse [001] direction. This causes the conduction band-edge to blue-shift with respect to the valence band. In addition, the minimum allowed energies in the bands are shifted with respect to their edges by quantization.

The first peak in the 250 \AA\ Ge/Ge$_{0.8}$Si$_{0.2}$ spectrum begins to rise at 0.84 eV. This is closer to the bulk Ge value of 0.798 eV because of less strain and larger well width as compared to the 100 \AA\ QW. The larger Ge content in the Ge$_{0.8}$Si$_{0.2}$ barriers results in less of lattice mismatch in the Ge well and so less of strain-induced blue-shift. Quantization becomes less significant in larger wells because of the $1/L_{z}^2$ dependence of the confinement energy for a square well.

It is evident that the two QWs generate very different spectra. The 100 \AA\ well has a steep rise, and then a series of successively weaker undulations appear before the signal falls to zero. The 250 \AA\ well on the other hand shows distinct peaks that oscillate through zero without any peak dominating over the rest. In both cases, the peaks broaden as photon energy increases. The samples being multi-layered, the influence of Fabry-Perot interference oscillations on the PR spectra had to be investigated.~\cite{huang} The inset in Fig.~\ref{fig:348ang_inset} shows PR from the 250 \AA\ sample, whose spectrum shows several oscillations, with the probe incident at $22^{\circ}$ and at $62^{\circ}$. Identical spectra which are not phase-shifted with respect to each other, are obtained. From this we conclude that interference effects do not determine the optical response of our samples.

\section{Simulation}
At a 2D critical point,~\cite{aspnes} the analytical line-shape expression for modulated reflectance at low fields is given by Eq.~(\ref{prfit}) with the exponent 2.5 replaced by 3. This has been widely used to characterize undoped QWs by PR. However in a QW of large $L_z$, the transition energies are closer spaced making it difficult to fit their PR with the Aspnes formula, even at low fields. At higher fields, spectra can be fitted by the FK expression, derived for bulk semiconductors.~\cite{shen} Therefore, to describe QW PR in high electric fields as a function of $L_z$, one has to perform an independent calculation. To this end, the quantized levels and wavefunctions in a QW under an electric field were calculated. They were then related to the optical absorption $\alpha$. These two steps were done with the help of a ready-made program~\footnote{http://www.wsi.tu-muenchen.de/nextnano3/} (freely downloadable). The absorption with and without the electric field (laser blocked or incident on the sample) was further processed by our own code to yield the PR signal $\Delta R/R$.

\begin{figure*}[hbt]
\includegraphics[width=12cm, angle=0]{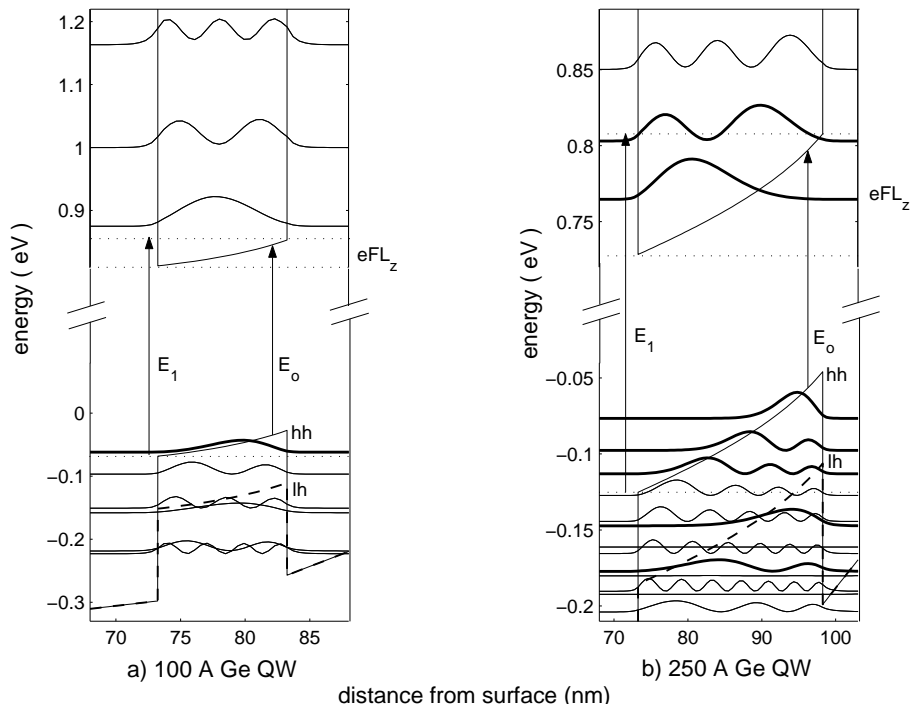}
\caption{\label{fig:wavefn}Square of the wavefunctions in a) the 100 \AA\ QW and b) the 250 \AA\ QW, as calculated by the program. The lowest valence band is the heavy-hole band, the dashed line indicates the higher energy strain-split light-hole band. The bold lines show the wavefunctions that contribute to the FK effect. Note that the scale used in a) and b) differs.}
\end{figure*}

Two checks were run on the downloaded program. It was used to solve the one-band Schroedinger equation under the effective-mass approximation to determine the quantized levels in the 100 \AA\ Ge/Ge$_{0.7}$Si$_{0.3}$ heterostructure. No electric field was considered, but the effect of strain was incorporated. It was ensured that the energy levels calculated matched those determined by solving for the energy levels in a finite square well graphically. Next the 8-band Schroedinger-Poisson equation was solved for the quantized levels and wavefunctions in a GaAs QW under an electric field of $10^5$ V/cm, for $L_z$ = 100 \AA\ and 300 \AA. The interband absorption was obtained from Eqs.~(\ref{eqn:dk}) and (\ref{da}). The absorption showed a trend from QCFK to FK behavior at $L_z$ = 300 \AA\, reproducing the results of Miller \textit{et al.}~\cite{miller}

Fig.~\ref{fig:wavefn} plots the square of the wavefunctions for both samples in the conduction and valence band, with the built-in field present. The valence band degeneracy is split by the strain into the heavy-hole band (solid-line) and the light-hole band (dashed line). We see that the 100 \AA\ Ge/Ge$_{0.7}$Si$_{0.3}$ sample is only weakly perturbed by the field (Fig.~\ref{fig:wavefn}(a)). The wavefunctions retain their zero-field sinusoidal behavior even though a field $F$ of average value 3.9$\times$10$^4$ V/cm is applied across the well. Since excitonic effects are not important, the effect of the electric field on the quantized levels and wavefunctions in the 100 \AA\ QW sample is referred to as the QCFKE.~\cite{millerd} In the 250 \AA\ Ge/Ge$_{0.8}$Si$_{0.2}$ sample, several wavefunctions tunnel into the forbidden gap (shown by bold lines in Fig.~\ref{fig:wavefn}(b)), assisted by the 3.0$\times$10$^4$ V/cm electric field. They correspond to kinetic energies less than $eFL_z$, the energy gained by the particle from the field $F$ by acceleration across a distance $L_z$. These wavefunctions have a pronounced Airy character, and lead to bulk-like FK oscillations for transition energies less than $E_1=E_o+eFL_z$, where $E_o$ is the relevant band-gap. Hence by examining the wavefunctions in the QW, one expects that the samples must show a trend towards bulk electro-absorption in their spectra.

For transition energies between $E_1$ and (not shown in Fig.~\ref{fig:wavefn}) $E_2=E_b-eFL_z$, $E_b$ is the barrier height, the QW wavefunctions are almost unperturbed by the electric field and return to their zero-field sinusoidal behavior. At still higher energies, the particles are unconfined and their energies are closely spaced.

\begin{table*}
\caption{Values of input parameters used in the calculation. $a$ is the lattice constant, $\epsilon$ is the static dielectric constant, $c_{11}$ and $c_{12}$ are the elastic constants, $E_o$ is the direct edge at the $\Gamma$-point, $\Delta_o$ is the spin-orbit splitting, $\gamma_1$, $\gamma_2$ and $\gamma_3$ are the valence band Luttinger parameters, $m_e$ is the conduction band $\Gamma$-point mass, $B$ is the inversion symmetry parameter, $b$ is the uniaxial (001) valence band deformation potential, and $dE_o/dP$ is the hydrostatic pressure induced shift of $E_o$. For the Ge$_y$Si$_{1-y}$ alloy, values were obtained by interpolating Ge and Si. Since the Ge$_y$Si$_{1-y}$ alloy barriers are unstrained, the strain parameters of Si were not used and are shown as blank entries. The relative band alignment between Ge and the alloy layers follows Ref.~\onlinecite{walle}. All parameters are from Ref.~\onlinecite{dargys} except as indicated. The optical matrix element $P$ (not shown in the table) was determined from the tabulated values using the equation:~\cite{people2} $m_o/m_e=1+2m_oP^2(2\Delta_o+3E_o)/3\hbar^2E_o(E_o+\Delta_o)$.}
\begin{ruledtabular}
\begin{tabular}{ccccccccccccccc}
 &$a$&$\epsilon$&$c_{11}$&$c_{12}$&$E_o$&$\Delta_o$&$\gamma_1$&$\gamma_2$&$\gamma_3$
 &$m_{e}$&$B$&$b$&$\frac{dE_o}{dP}$ \\ 
 &(\AA)& &(GPa)&(GPa)&(eV)&(eV)& & & &(m$_o$)& &(eV)&(eV/GPa)\\ \colrule
Ge&5.65&15.94&128.5&48.3&0.798&0.289&13.38&4.24&5.69&0.038&0&-2.55\footnotemark[1]&0.121 \\
Si&5.43&11.9&-&-&4.00\footnotemark[2]&0.044\footnotemark[3]&4.285\footnotemark[3]&0.339\footnotemark[3]&1.446\footnotemark[3]&0.64\footnotemark[4]&0&-&-\\
\end{tabular}
\end{ruledtabular}
\footnotetext[1]{Ref.~\onlinecite{walle}.}
\footnotetext[2]{Ref.~\onlinecite{kline}.}
\footnotetext[3]{Ref.~\onlinecite{cardona}.}
\footnotetext[4]{Ref.~\onlinecite{people1}.}
 \label{table1}
 \end{table*}

Table \ref{table1}. indicates the values of all the physical quantities used in calculating the absorption $\alpha$. To calculate the reflectance $R$, measured $n(\omega)$ and $k(\omega)$ values of bulk Ge~\cite{palik1} were used for the QW. Measured values for a relaxed Ge$_{0.75}$Si$_{0.25}$ alloy~\cite{palik2} were used for the barrier layers. Identical simulation parameters were used for both samples with no adjustment. Details about the calculation are provided in the appendix.

\begin{figure}[bht]
\includegraphics[width=8cm, angle=0]{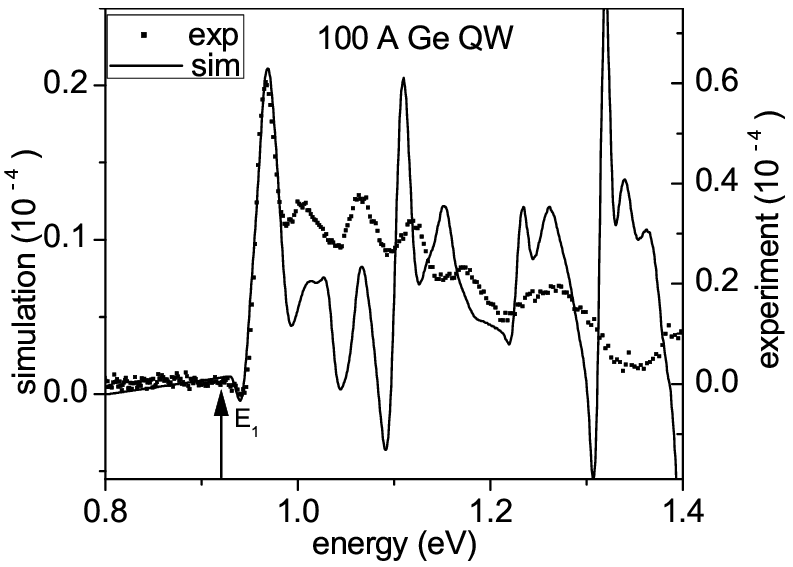}
\caption{\label{fig:fit411}Fit of 100 \AA\ QW. The scales have not been adjusted except for a vertical shift to subtract the experimental background.}
\end{figure}

\section{Simulation results}

\begin{figure}[bhp]
\includegraphics[width=8cm, angle=0]{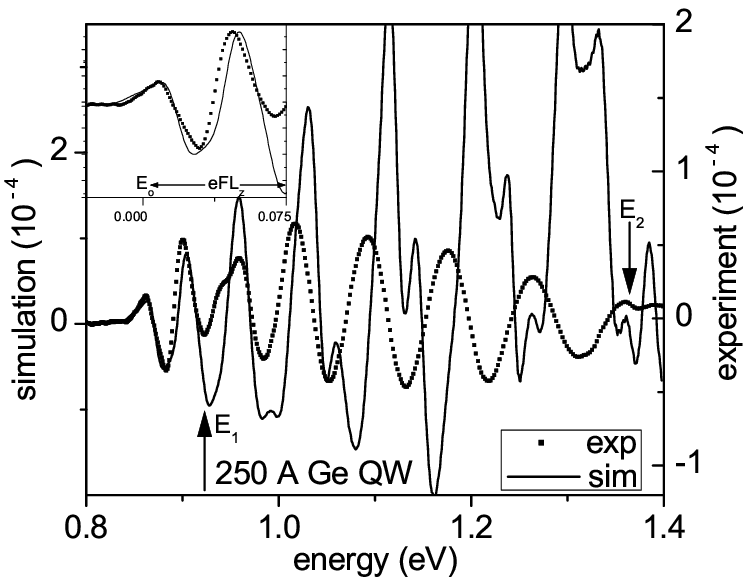}
\caption{\label{fig:fit348}Fit of 250 \AA\ QW. The scales are unadjusted except for a vertical shift to null experimental background. The inset shows an expanded view close to the $E_o$ gap}
\end{figure}

To verify the effect of $L_z$ in determining the difference in PR from the two QWs, having comparable electric fields, we plot the simulated along with the experimental spectra in Figs.~\ref{fig:fit411} and~\ref{fig:fit348}. Setting $E_o$ equal to the heavy-hole band-gap, the maximum energy $E_1$ upto which FK oscillations can occur has been depicted in the figures by an arrow. Both in simulation and experiment, we find that while the 100 \AA\ QW is featureless upto $E_1$, the 250 \AA\ QW shows two peaks. The inset in Fig.~\ref{fig:fit348} shows the PR from the 250 \AA\ well in the region close to the gap, with the abscissa plotted relative to $E_o$. A FK tail is also visible below the gap, caused by the overlap of those parts of the Airy functions that exponentially decay into it.

The remaining part of the spectrum lying between $E_1$ and $E_2$ involves at least one wavefunction that is not an Airy function. $E_2=1.72$ eV for the 100 \AA\ well (lying outside the experimental range) and $E_2=1.365$ eV for the 250 \AA\ well (shown by an arrow in Fig.~\ref{fig:fit348}). The first three peaks match experiment in Fig.~\ref{fig:fit411}, but the agreement is poorer in Fig.~\ref{fig:fit348}, especially at higher energies.

The most likely reason for the mismatch at higher energies can be seen from Fig.~\ref{fig:calcband348}. Because of the form of the sample potential, the largest modulation takes place at the surface. For the computation to be practically feasible, the wavefunctions were simulated over a region encompassing the QW and extending by a distance equal to $L_z$, in either direction. Since the surface region was not included in the simulation, wavefunctions calculated near the mouth of the QW were not quite correct, leading to errors in the spectra at higher energies. However wavefunctions close to the bottom of the QW are unaffected by the surface potential. This explains why the calculated and experimental spectra are in good agreement in the vicinity of the band-edge.

\section{Conclusion}
The PR spectra obtained from the 100 \AA\ and 250 \AA\ QWs were very unlike each other. Since the samples had different $L_z$, it was the most likely cause for the disparate spectra. This was confirmed by a simulation: the 250 \AA\ QW had several Airy-like wavefunctions tunnelling into the band-gap, similar to the bulk FKE. This clearly indicated that PR from the 100 \AA\ and the 250 \AA\ Ge QWs displayed a transition from the QCFKE to the FKE.

For this transition, a critical well-width $L_z^{(c)}$ = 300 \AA\ had been predicted~\cite{miller} for GaAs ($m_{hh}$ = 0.34$m_o$, $m_e$ = 0.0665$m_o$), in a field $F$ = $10^5$ V/cm. To facilitate comparison with our experiment in a 250 \AA\ Ge QW ($m_{hh}$ = $m_o/(\gamma_1-2\gamma_2)$ = 0.204$m_o$, $m_e$ = 0.038$m_o$, $F$ = 3$\times$10$^4$ V/cm), an approximate scaling relation for $L_z^{(c)}$ can be obtained by equating the energy of a particle-in-a-box, $h^2/8\mu L_z^2$, with the energy gained from the field, $eFL_z$
\begin{equation}
L_z^{(c)}\approx\left(\frac{h^2}{8\mu eF}\right)^{1/3}, \label{eq:Lz}
\end{equation}
where $\mu$ is the reduced mass for the heavy-hole and the conduction band. From Eqn.~(\ref{eq:Lz}), the two-band infinite well model of Ref.~\onlinecite{miller} predicts $L_z^{(c)}$ = 540 \AA\ in Ge. For a finite barrier, it is expected that the transition should occur at a somewhat smaller $L_z^{(c)}$, since the quantized levels are closer spaced and in an electric field more levels can tunnel into the band-gap. Consequently our result for a 250 \AA\ Ge finite QW, based on a realistic description of the band structure, is found to be consistent with the theoretical estimate.

\begin{acknowledgments}
We wish to thank Prof. Y. H. Xie (UCLA) for the Ge QW samples. We are also grateful to the developers of the program for making it freely available.
\end{acknowledgments}

\appendix*
\section{Details of the calculation}
\subsection{Modulated absorption}
The 8-band Schroedinger-Poisson equation for the structure was first solved so as to obtain the quantized energy levels and their wavefunctions at the $\Gamma$-point in the conduction and valence band. For each quantized level, the energy eigenvalues and wavefunctions were further obtained over a 2D grid of k-points in the Brillouin zone. The \textbf{k.p} coupling between the conduction, heavy-hole, light-hole and split-off-hole bands ensures that non-parabolicity and anisotropy of the E-\textbf{k} dispersion in the plane of the QW are included in the calculation. It is important to include the effect of non-parabolicity since the well depth was comparatively large (645 meV in the conduction band of the 100 \AA\ QW). The calculation also accounted for the modification of the $\Gamma$-point masses by strain present in the samples.

The imaginary part of the optical susceptibility $\chi^{(2)}$ was then evaluated in the Ge QW by
\begin{widetext}
\begin{equation}
\chi^{(2)}_{ij}(\omega)=\left(\frac{1}{4\pi}\right)\frac{\pi e^2}{m_o\omega^2\epsilon_o}\frac{2}{\Omega}\sum_{c,v,k_{\parallel}}(f_{vk_{\parallel}}-f_{ck_{\parallel}}) 
  \langle ck_{\parallel}|\nabla_i|vk_{\parallel}\rangle\langle
vk_{\parallel}|\nabla_j|ck_{\parallel}\rangle\delta(E_{ck_{\parallel}}-E_{vk_{\parallel}}-\hbar\omega) ,
\end{equation}
\end{widetext}
where $c$ and $v$ label the quantized energies in the conduction and valence band, $k_{\parallel}$ labels the in-plane $k$-vectors, $f$ is the Fermi distribution function, $m_o$ is the free electron mass and $\Omega$ is the normalization volume. From $\chi^{(2)}(\omega)$, we can obtain the imaginary part of the dielectric constant $\epsilon^{(2)}(\omega)$  which in turn can be related to the imaginary part of the refractive index $k$. Modulation by the laser modifies the quantized energies in the QW, changing the imaginary susceptibility and the refractive index by $\Delta\chi^{(2)}$ and $\Delta k$
\begin{eqnarray}
\Delta k(\omega) &=& \frac{\Delta\epsilon^{(2)}(\omega)}{2n(\omega)} \nonumber\\ \label{eqn:neglectdn}
                 &=& \frac{4\pi( \beta^*_i\Delta\chi^{(2)}_{ij}\beta_j)}{2n(\omega)},\label{eqn:dk}
\end{eqnarray}
where $\beta_i$ is the polarization component of the probe and $n$ is the real part of the refractive index. The fractional change in the real part of the refractive index $\Delta n$ has been neglected in comparison with $n$ in writing Eqn.~\ref{eqn:neglectdn}. $\Delta k$ can also be related to the change in the absorption coefficient $\alpha$ by
\begin{equation}
\Delta\alpha(\omega)=\frac{2\omega}{c}\Delta k(\omega). \label{da}
\end{equation}

\subsection{Photoreflectance signal}
Since the modulated reflectance $\Delta R$ depends on both $\Delta n$ and $\Delta k$, we determine $\Delta n(\omega)$ by a Kramers-Kronig transform of $\Delta\alpha$ (Eq.~(\ref{da}))
\begin{equation}
\Delta n(\omega)=\frac{c}{\pi}\mathcal{P}\int_0^\infty\frac{\Delta\alpha(\omega')d\omega'}{\omega'^2-\omega^2}.
\end{equation}
The difference quantity $\Delta\alpha$ is localized in frequency and the integral converges rapidly. To determine the $\Delta R/R$ spectra of our multilayered samples,~\cite{huang} consider two reflecting interfaces first. For mutually perpendicular polarizations ($p$ and $s$), the complex ratio of reflected to incident light is
\begin{equation}
r^{p,s}=\frac{r^{p,s}_{1,2}+r^{p,s}_{2,3}e^{-ix_2}}{1+r^{p,s}_{1,2}r^{p,s}_{2,3}e^{-ix_2}}, \label{refl}
\end{equation}
where $r^{p,s}_{i,i+1}$ is the reflectivity at the interface between media $i$ and $i+1$, $x_2=(2\pi/\lambda)2n_2 d_2 cos\theta_2$ is the phase difference between reflected beams 1 and 2, and $\lambda$ is the wavelength of the probe light. The reflectivity can be calculated from~\cite{cardona}
\begin{eqnarray}
r^{p}_{i,i+1}&=&\frac{\frac{N_{i+1}^2}{N_i^2}cos\theta_i-\sqrt{\frac{N_{i+1}^2}{N_i^2}-sin^2\theta_i}}{\frac{N_{i+1}^2}{N_i^2}cos\theta_i+\sqrt{\frac{N_{i+1}^2}{N_i^2}-sin^2\theta_i}}, \label{r1}\\
r^{s}_{i,i+1}&=&\frac{cos\theta_i-\sqrt{\frac{N_{i+1}^2}{N_i^2}-sin^2\theta_i}}{cos\theta_i+\sqrt{\frac{N_{i+1}^2}{N_i^2}-sin^2\theta_i}}, \label{r2}
\end{eqnarray}
where $N_i=n_i+ik_i$, and $\theta_i$ is the incident angle, in media $i$. We considered three interfaces wherein $r^{p,s}_{2,3}$ is determined by Eqs.~(\ref{refl})-(\ref{r2}) and plugged back into Eq.~(\ref{refl}) to obtain $r^{p,s}$ as a function of $r^{p,s}_{1,2}$, $r^{p,s}_{2,3}$ and $r^{p,s}_{3,4}$. The reflectance of the sample for unpolarized light is then given by
\begin{equation} R=\frac{1}{2}(r^p r^{p^*} + r^s r^{s^*}). \end{equation}
Finally the modulated reflectance is determined from
\begin{eqnarray}
\Delta R&=&R(n_{Ge},n_{alloy},k_{Ge},k_{alloy},d_i,\Delta n_{Ge},\Delta k_{Ge},\omega)\nonumber\\
        & &-R(n_{Ge},n_{alloy},k_{Ge},k_{alloy},d_i,\omega),
\end{eqnarray}
where $d_i$ corresponds to the thickness of the various layers in the sample and $n$ and $k$ in the Ge and alloy layers are functions of $\omega$. Normalizing to $R$ gives the PR signal $\Delta R/R$.

\bibliography{arjun}

\begin{thebibliography}{31}
\expandafter\ifx\csname natexlab\endcsname\relax\def\natexlab#1{#1}\fi
\expandafter\ifx\csname bibnamefont\endcsname\relax
  \def\bibnamefont#1{#1}\fi
\expandafter\ifx\csname bibfnamefont\endcsname\relax
  \def\bibfnamefont#1{#1}\fi
\expandafter\ifx\csname citenamefont\endcsname\relax
  \def\citenamefont#1{#1}\fi
\expandafter\ifx\csname url\endcsname\relax
  \def\url#1{\texttt{#1}}\fi
\expandafter\ifx\csname urlprefix\endcsname\relax\def\urlprefix{URL }\fi
\providecommand{\bibinfo}[2]{#2}
\providecommand{\eprint}[2][]{\url{#2}}

\bibitem[{\citenamefont{Schmeller et~al.}(1994)\citenamefont{Schmeller, Hansen,
  Kotthaus, Tr{\"{a}}nkle, and Weimann}}]{schmeller}
\bibinfo{author}{\bibfnamefont{A.}~\bibnamefont{Schmeller}},
  \bibinfo{author}{\bibfnamefont{W.}~\bibnamefont{Hansen}},
  \bibinfo{author}{\bibfnamefont{J.~P.} \bibnamefont{Kotthaus}},
  \bibinfo{author}{\bibfnamefont{G.}~\bibnamefont{Tr{\"{a}}nkle}},
  \bibnamefont{and} \bibinfo{author}{\bibfnamefont{G.}~\bibnamefont{Weimann}},
  \bibinfo{journal}{Appl.\ Phys. Lett.} \textbf{\bibinfo{volume}{64}},
  \bibinfo{pages}{330} (\bibinfo{year}{1994}).

\bibitem[{\citenamefont{Miller et~al.}(1986{\natexlab{a}})\citenamefont{Miller,
  Weiner, and Chemla}}]{millerc}
\bibinfo{author}{\bibfnamefont{D.~A.~B.} \bibnamefont{Miller}},
  \bibinfo{author}{\bibfnamefont{J.~S.} \bibnamefont{Weiner}},
  \bibnamefont{and} \bibinfo{author}{\bibfnamefont{D.~S.}
  \bibnamefont{Chemla}}, \bibinfo{journal}{IEEE J. Quantum Electron.}
  \textbf{\bibinfo{volume}{22}}, \bibinfo{pages}{1816}
  (\bibinfo{year}{1986}{\natexlab{a}}).

\bibitem[{\citenamefont{Merbach et~al.}(1998)\citenamefont{Merbach,
  Sch{\"{o}}ll, Ebeling, Michler, and Gutowski}}]{merbach}
\bibinfo{author}{\bibfnamefont{D.}~\bibnamefont{Merbach}},
  \bibinfo{author}{\bibfnamefont{E.}~\bibnamefont{Sch{\"{o}}ll}},
  \bibinfo{author}{\bibfnamefont{W.}~\bibnamefont{Ebeling}},
  \bibinfo{author}{\bibfnamefont{P.}~\bibnamefont{Michler}}, \bibnamefont{and}
  \bibinfo{author}{\bibfnamefont{J.}~\bibnamefont{Gutowski}},
  \bibinfo{journal}{Phys.\ Rev. B} \textbf{\bibinfo{volume}{58}},
  \bibinfo{pages}{10709} (\bibinfo{year}{1998}).

\bibitem[{\citenamefont{Miller et~al.}(1986{\natexlab{b}})\citenamefont{Miller,
  Chemla, and Schmitt-Rink}}]{miller}
\bibinfo{author}{\bibfnamefont{D.~A.~B.} \bibnamefont{Miller}},
  \bibinfo{author}{\bibfnamefont{D.~S.} \bibnamefont{Chemla}},
  \bibnamefont{and}
  \bibinfo{author}{\bibfnamefont{S.}~\bibnamefont{Schmitt-Rink}},
  \bibinfo{journal}{Phys.\ Rev. B} \textbf{\bibinfo{volume}{33}},
  \bibinfo{pages}{6976} (\bibinfo{year}{1986}{\natexlab{b}}).

\bibitem[{\citenamefont{Schmitt-Rink et~al.}(1989)\citenamefont{Schmitt-Rink,
  Chemla, and Miller}}]{millerd}
\bibinfo{author}{\bibfnamefont{S.}~\bibnamefont{Schmitt-Rink}},
  \bibinfo{author}{\bibfnamefont{D.~S.} \bibnamefont{Chemla}},
  \bibnamefont{and} \bibinfo{author}{\bibfnamefont{D.~A.~B.}
  \bibnamefont{Miller}}, \bibinfo{journal}{Adv. Phys}
  \textbf{\bibinfo{volume}{38}}, \bibinfo{pages}{89} (\bibinfo{year}{1989}).

\bibitem[{\citenamefont{Tharmalingam}(1963)}]{tharma}
\bibinfo{author}{\bibfnamefont{K.}~\bibnamefont{Tharmalingam}},
  \bibinfo{journal}{Phys. Rev.} \textbf{\bibinfo{volume}{130}},
  \bibinfo{pages}{2204} (\bibinfo{year}{1963}).

\bibitem[{\citenamefont{Aspnes}(1966)}]{aspnes1}
\bibinfo{author}{\bibfnamefont{D.~E.} \bibnamefont{Aspnes}},
  \bibinfo{journal}{Phys. Rev.} \textbf{\bibinfo{volume}{147}},
  \bibinfo{pages}{554} (\bibinfo{year}{1966}).

\bibitem[{\citenamefont{Wang et~al.}(1995)\citenamefont{Wang, Chen, Kuan, Shei,
  and Su}}]{wang}
\bibinfo{author}{\bibfnamefont{D.~P.} \bibnamefont{Wang}},
  \bibinfo{author}{\bibfnamefont{C.~T.} \bibnamefont{Chen}},
  \bibinfo{author}{\bibfnamefont{H.}~\bibnamefont{Kuan}},
  \bibinfo{author}{\bibfnamefont{S.~C.} \bibnamefont{Shei}}, \bibnamefont{and}
  \bibinfo{author}{\bibfnamefont{Y.~K.} \bibnamefont{Su}}, \bibinfo{journal}{J.
  Appl. Phys.} \textbf{\bibinfo{volume}{78}}, \bibinfo{pages}{2117}
  (\bibinfo{year}{1995}).

\bibitem[{\citenamefont{Collins et~al.}(1986)\citenamefont{Collins,
  v.~Klitzing, and Ploog}}]{collins}
\bibinfo{author}{\bibfnamefont{R.~T.} \bibnamefont{Collins}},
  \bibinfo{author}{\bibfnamefont{K.}~\bibnamefont{v.~Klitzing}},
  \bibnamefont{and} \bibinfo{author}{\bibfnamefont{K.}~\bibnamefont{Ploog}},
  \bibinfo{journal}{Phys.\ Rev. B.} \textbf{\bibinfo{volume}{33}},
  \bibinfo{pages}{4378} (\bibinfo{year}{1986}).

\bibitem[{\citenamefont{Yamanaka et~al.}(1986)\citenamefont{Yamanaka, Fukunaga,
  Tsukada, Kobayashi, and Ishii}}]{yamanaka}
\bibinfo{author}{\bibfnamefont{K.}~\bibnamefont{Yamanaka}},
  \bibinfo{author}{\bibfnamefont{T.}~\bibnamefont{Fukunaga}},
  \bibinfo{author}{\bibfnamefont{N.}~\bibnamefont{Tsukada}},
  \bibinfo{author}{\bibfnamefont{K.}~\bibnamefont{Kobayashi}},
  \bibnamefont{and} \bibinfo{author}{\bibfnamefont{M.}~\bibnamefont{Ishii}},
  \bibinfo{journal}{Appl. Phys. Lett.} \textbf{\bibinfo{volume}{48}},
  \bibinfo{pages}{840} (\bibinfo{year}{1986}).

\bibitem[{\citenamefont{Chen and Lu}(2000)}]{chen}
\bibinfo{author}{\bibfnamefont{R.~B.} \bibnamefont{Chen}} \bibnamefont{and}
  \bibinfo{author}{\bibfnamefont{Y.}~\bibnamefont{Lu}}, \bibinfo{journal}{Solid
  State Commun.} \textbf{\bibinfo{volume}{114}}, \bibinfo{pages}{117}
  (\bibinfo{year}{2000}).

\bibitem[{\citenamefont{Keil et~al.}(1991)\citenamefont{Keil, Linder, Schmidt,
  D{\"{o}}hler, and Miller}}]{dohler}
\bibinfo{author}{\bibfnamefont{U.~D.} \bibnamefont{Keil}},
  \bibinfo{author}{\bibfnamefont{N.}~\bibnamefont{Linder}},
  \bibinfo{author}{\bibfnamefont{K.}~\bibnamefont{Schmidt}},
  \bibinfo{author}{\bibfnamefont{G.}~\bibnamefont{D{\"{o}}hler}},
  \bibnamefont{and} \bibinfo{author}{\bibfnamefont{J.~N.}
  \bibnamefont{Miller}}, \bibinfo{journal}{Phys.\ Rev. B}
  \textbf{\bibinfo{volume}{44}}, \bibinfo{pages}{13504} (\bibinfo{year}{1991}).

\bibitem[{\citenamefont{Sanders and Bajaj}(1987)}]{sanders}
\bibinfo{author}{\bibfnamefont{G.~D.} \bibnamefont{Sanders}} \bibnamefont{and}
  \bibinfo{author}{\bibfnamefont{K.~K.} \bibnamefont{Bajaj}},
  \bibinfo{journal}{Phys.\ Rev. B} \textbf{\bibinfo{volume}{35}},
  \bibinfo{pages}{2308} (\bibinfo{year}{1987}).

\bibitem[{\citenamefont{Pearsall et~al.}(1986)\citenamefont{Pearsall, Pollak,
  Bean, and Hull}}]{pearsall}
\bibinfo{author}{\bibfnamefont{T.~P.} \bibnamefont{Pearsall}},
  \bibinfo{author}{\bibfnamefont{F.~H.} \bibnamefont{Pollak}},
  \bibinfo{author}{\bibfnamefont{J.~C.} \bibnamefont{Bean}}, \bibnamefont{and}
  \bibinfo{author}{\bibfnamefont{R.}~\bibnamefont{Hull}},
  \bibinfo{journal}{Phys.\ Rev. B} \textbf{\bibinfo{volume}{33}},
  \bibinfo{pages}{6821} (\bibinfo{year}{1986}).

\bibitem[{\citenamefont{Rodrigues et~al.}(1992)\citenamefont{Rodrigues,
  Cerdeira, and Bean}}]{rodrigues}
\bibinfo{author}{\bibfnamefont{P.~A.~M.} \bibnamefont{Rodrigues}},
  \bibinfo{author}{\bibfnamefont{F.}~\bibnamefont{Cerdeira}}, \bibnamefont{and}
  \bibinfo{author}{\bibfnamefont{J.~C.} \bibnamefont{Bean}},
  \bibinfo{journal}{Phys.\ Rev. B} \textbf{\bibinfo{volume}{46}},
  \bibinfo{pages}{15263} (\bibinfo{year}{1992}).

\bibitem[{\citenamefont{Yaguchi et~al.}(1994)\citenamefont{Yaguchi, Tai,
  Takemasa, Onabe, Ito, and Shiraki}}]{yaguchi}
\bibinfo{author}{\bibfnamefont{H.}~\bibnamefont{Yaguchi}},
  \bibinfo{author}{\bibfnamefont{K.}~\bibnamefont{Tai}},
  \bibinfo{author}{\bibfnamefont{K.}~\bibnamefont{Takemasa}},
  \bibinfo{author}{\bibfnamefont{K.}~\bibnamefont{Onabe}},
  \bibinfo{author}{\bibfnamefont{R.}~\bibnamefont{Ito}}, \bibnamefont{and}
  \bibinfo{author}{\bibfnamefont{Y.}~\bibnamefont{Shiraki}},
  \bibinfo{journal}{Phys.\ Rev. B} \textbf{\bibinfo{volume}{49}},
  \bibinfo{pages}{7394} (\bibinfo{year}{1994}).

\bibitem[{\citenamefont{Xie et~al.}(1993{\natexlab{a}})\citenamefont{Xie,
  Monroe, Fitzgerald, Silverman, Thiel, and Watson}}]{xie1}
\bibinfo{author}{\bibfnamefont{Y.~H.} \bibnamefont{Xie}},
  \bibinfo{author}{\bibfnamefont{D.}~\bibnamefont{Monroe}},
  \bibinfo{author}{\bibfnamefont{E.~A.} \bibnamefont{Fitzgerald}},
  \bibinfo{author}{\bibfnamefont{P.~J.} \bibnamefont{Silverman}},
  \bibinfo{author}{\bibfnamefont{F.~A.} \bibnamefont{Thiel}}, \bibnamefont{and}
  \bibinfo{author}{\bibfnamefont{G.~P.} \bibnamefont{Watson}},
  \bibinfo{journal}{Appl. Phys. Lett.} \textbf{\bibinfo{volume}{63}},
  \bibinfo{pages}{2263} (\bibinfo{year}{1993}{\natexlab{a}}).

\bibitem[{\citenamefont{Xie et~al.}(1993{\natexlab{b}})\citenamefont{Xie,
  Fitzgerald, Monroe, Silverman, and Watson}}]{xie2}
\bibinfo{author}{\bibfnamefont{Y.~H.} \bibnamefont{Xie}},
  \bibinfo{author}{\bibfnamefont{E.~A.} \bibnamefont{Fitzgerald}},
  \bibinfo{author}{\bibfnamefont{D.}~\bibnamefont{Monroe}},
  \bibinfo{author}{\bibfnamefont{P.~J.} \bibnamefont{Silverman}},
  \bibnamefont{and} \bibinfo{author}{\bibfnamefont{G.~P.}
  \bibnamefont{Watson}}, \bibinfo{journal}{J. Appl. Phys.}
  \textbf{\bibinfo{volume}{73}}, \bibinfo{pages}{8364}
  (\bibinfo{year}{1993}{\natexlab{b}}).

\bibitem[{\citenamefont{Madhavi et~al.}(2001)\citenamefont{Madhavi,
  Venkataraman, and Xie}}]{madhavijap}
\bibinfo{author}{\bibfnamefont{S.}~\bibnamefont{Madhavi}},
  \bibinfo{author}{\bibfnamefont{V.}~\bibnamefont{Venkataraman}},
  \bibnamefont{and} \bibinfo{author}{\bibfnamefont{Y.~H.} \bibnamefont{Xie}},
  \bibinfo{journal}{J. Appl. Phys} \textbf{\bibinfo{volume}{89}},
  \bibinfo{pages}{2497} (\bibinfo{year}{2001}).

\bibitem[{\citenamefont{Madhavi et~al.}(2000)\citenamefont{Madhavi,
  Venkataraman, Sturm, and Xie}}]{madhaviprb}
\bibinfo{author}{\bibfnamefont{S.}~\bibnamefont{Madhavi}},
  \bibinfo{author}{\bibfnamefont{V.}~\bibnamefont{Venkataraman}},
  \bibinfo{author}{\bibfnamefont{J.~C.} \bibnamefont{Sturm}}, \bibnamefont{and}
  \bibinfo{author}{\bibfnamefont{Y.~H.} \bibnamefont{Xie}},
  \bibinfo{journal}{Phys. Rev. B} \textbf{\bibinfo{volume}{61}},
  \bibinfo{pages}{16807} (\bibinfo{year}{2000}).

\bibitem[{\citenamefont{Aspnes}(1973)}]{aspnes}
\bibinfo{author}{\bibfnamefont{D.~E.} \bibnamefont{Aspnes}},
  \bibinfo{journal}{Surf. Sci.} \textbf{\bibinfo{volume}{37}},
  \bibinfo{pages}{418} (\bibinfo{year}{1973}).

\bibitem[{\citenamefont{Huang et~al.}(1989)\citenamefont{Huang, Mui, and
  Morko\c{c}}}]{huang}
\bibinfo{author}{\bibfnamefont{D.}~\bibnamefont{Huang}},
  \bibinfo{author}{\bibfnamefont{D.}~\bibnamefont{Mui}}, \bibnamefont{and}
  \bibinfo{author}{\bibfnamefont{H.}~\bibnamefont{Morko\c{c}}},
  \bibinfo{journal}{J. Appl. Phys.} \textbf{\bibinfo{volume}{66}},
  \bibinfo{pages}{358} (\bibinfo{year}{1989}).

\bibitem[{\citenamefont{Shen and Pollak}(1990)}]{shen}
\bibinfo{author}{\bibfnamefont{H.}~\bibnamefont{Shen}} \bibnamefont{and}
  \bibinfo{author}{\bibfnamefont{F.~H.} \bibnamefont{Pollak}},
  \bibinfo{journal}{Phys. Rev. B} \textbf{\bibinfo{volume}{42}},
  \bibinfo{pages}{7097} (\bibinfo{year}{1990}).

\bibitem[{\citenamefont{de~Walle and Martin}(1986)}]{walle}
\bibinfo{author}{\bibfnamefont{C.~G.~V.} \bibnamefont{de~Walle}}
  \bibnamefont{and} \bibinfo{author}{\bibfnamefont{R.~M.}
  \bibnamefont{Martin}}, \bibinfo{journal}{Phys.\ Rev. B}
  \textbf{\bibinfo{volume}{34}}, \bibinfo{pages}{5621} (\bibinfo{year}{1986}).

\bibitem[{\citenamefont{Dargys and Kundrotas}(1994)}]{dargys}
\bibinfo{author}{\bibfnamefont{A.}~\bibnamefont{Dargys}} \bibnamefont{and}
  \bibinfo{author}{\bibfnamefont{J.}~\bibnamefont{Kundrotas}},
  \emph{\bibinfo{title}{Handbook on Physical Properties of Ge, Si, GaAs and
  InP}} (\bibinfo{publisher}{Science and Encyclopedia Publishers},
  \bibinfo{address}{Vilnius}, \bibinfo{year}{1994}).

\bibitem[{\citenamefont{People and Sputz}(1990)}]{people2}
\bibinfo{author}{\bibfnamefont{R.}~\bibnamefont{People}} \bibnamefont{and}
  \bibinfo{author}{\bibfnamefont{S.~K.} \bibnamefont{Sputz}},
  \bibinfo{journal}{Phys. Rev. B} \textbf{\bibinfo{volume}{41}},
  \bibinfo{pages}{8431} (\bibinfo{year}{1990}).

\bibitem[{\citenamefont{Kline et~al.}(1968)\citenamefont{Kline, Pollak, and
  Cardona}}]{kline}
\bibinfo{author}{\bibfnamefont{J.~S.} \bibnamefont{Kline}},
  \bibinfo{author}{\bibfnamefont{F.~H.} \bibnamefont{Pollak}},
  \bibnamefont{and} \bibinfo{author}{\bibfnamefont{M.}~\bibnamefont{Cardona}},
  \bibinfo{journal}{Helv. Phys. Acta} \textbf{\bibinfo{volume}{41}},
  \bibinfo{pages}{968} (\bibinfo{year}{1968}).

\bibitem[{\citenamefont{Yu and Cardona}(2001)}]{cardona}
\bibinfo{author}{\bibfnamefont{P.~Y.} \bibnamefont{Yu}} \bibnamefont{and}
  \bibinfo{author}{\bibfnamefont{M.}~\bibnamefont{Cardona}},
  \emph{\bibinfo{title}{Fundamentals of Semiconductors}}
  (\bibinfo{publisher}{Springer-Verlag}, \bibinfo{address}{Berlin},
  \bibinfo{year}{2001}).

\bibitem[{\citenamefont{People and Jackson}(1987)}]{people1}
\bibinfo{author}{\bibfnamefont{R.}~\bibnamefont{People}} \bibnamefont{and}
  \bibinfo{author}{\bibfnamefont{S.~A.} \bibnamefont{Jackson}},
  \bibinfo{journal}{Phys.\ Rev. B} \textbf{\bibinfo{volume}{36}},
  \bibinfo{pages}{1310} (\bibinfo{year}{1987}).

\bibitem[{\citenamefont{Potter}(1985)}]{palik1}
\bibinfo{author}{\bibfnamefont{R.~F.} \bibnamefont{Potter}}, in
  \emph{\bibinfo{booktitle}{Handbook of optical constants of Solids}}, edited
  by \bibinfo{editor}{\bibfnamefont{E.~D.} \bibnamefont{Palik}}
  (\bibinfo{publisher}{Academic Press}, \bibinfo{address}{London},
  \bibinfo{year}{1985}).

\bibitem[{\citenamefont{Huml\'{\i}\u{c}ek
  et~al.}(1998)\citenamefont{Huml\'{\i}\u{c}ek, Luke\u{s}, and
  Schmidt}}]{palik2}
\bibinfo{author}{\bibfnamefont{J.}~\bibnamefont{Huml\'{\i}\u{c}ek}},
  \bibinfo{author}{\bibfnamefont{F.}~\bibnamefont{Luke\u{s}}},
  \bibnamefont{and} \bibinfo{author}{\bibfnamefont{E.}~\bibnamefont{Schmidt}},
  in \emph{\bibinfo{booktitle}{Handbook of optical constants of Solids}},
  edited by \bibinfo{editor}{\bibfnamefont{E.~D.} \bibnamefont{Palik}}
  (\bibinfo{publisher}{Academic Press}, \bibinfo{address}{San Diego},
  \bibinfo{year}{1998}), vol.~\bibinfo{volume}{II}.

\end{thebibliography}

\end{document}